# Validated Modeling of Atmospheric-Pressure Anodic Arc

A. Khrabry[1], I. D. Kaganovich, A. Khodak, V. Vekselman, Y. Raitses

*Princeton Plasma Physics Laboratory, Princeton NJ 08542*

**Abstract:** We performed self-consistent modeling of the atmospheric-pressure plasmas produced by arc discharges. Special numerical procedure for coupling of the plasma current, emission current, ion current, sheath voltage drop, heat fluxes at plasma-electrode interfaces was developed and implemented into the ANSYS CFX code. Validation of the simulation results is performed through comparison of simulation results with the previously available experimental data obtained by a range of various plasma diagnostics.

1. **Basic equations for the arc model**

Modeling of the arc for nanomaterial synthesis and other applications should be able to predict plasma parameters for different arc configurations and operating conditions. We developed a model that can reliably predict arc properties and incorporated it into a numeral code. Previous carbon arc simulations[1] were limited due to use of simplifying assumptions, such as the local thermodynamic equilibrium (LTE) (the same temperature for the electrons and heavy species), and the ionization and recombination local equality (the Saha equation), as well as not accounting for the heat losses in ablation processes and heat generation during depositions, and not fully taking into account space-charge sheathes near electrodes, etc. All these processes are important for the short arcs and are accurately taken into account in our simulations[2].

In the previous studies, we developed one-dimensional self-consistent analytical[3] and numerical[4] models of the near-electrode regions and the arc column combined into a unified self-consistent model of the whole arc. Non-equilibrium processes in the plasma and space-charge sheath effects are taken into account, and coupled to heat transfer in the electrodes as shown in Fig.1. The analytical arc model is capable to predict the electric field, plasma, and gas density and temperature profiles in different arc regions, their sizes, and heat fluxes to the electrodes[3], as shown in Fig.2. This model aids in better understanding of cathodic and anodic processes and easy assessments of heat fluxes, voltages, temperature values, etc. For benchmarking of the arc model, in particular sheath model and plasma transport coefficients, a 1D code that resolves sheath regions was written and simulations were performed to compare with results of previous numerical studies by Almeida et al[5], and complete agreement was achieved[4], see Fig.1.

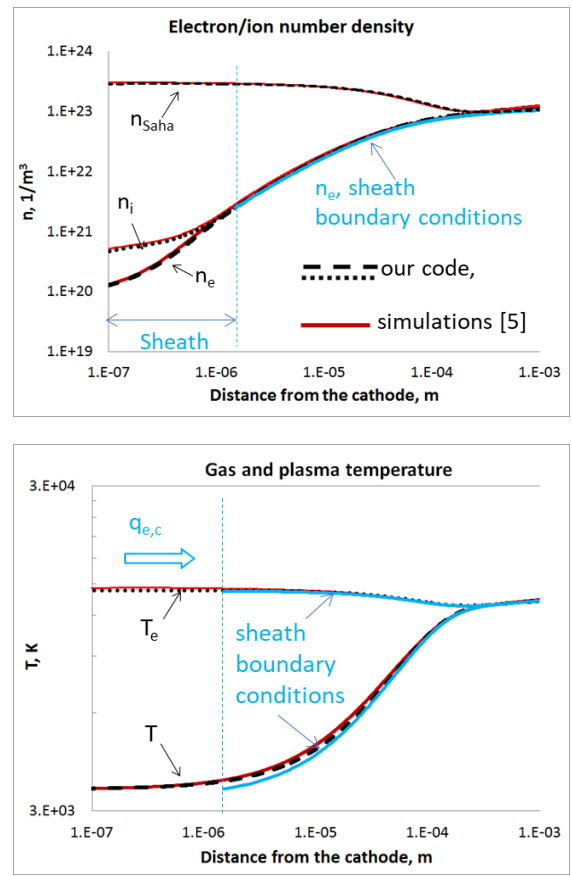

*Fig.1 Cross-verification of space-charge sheath models showing that sheath resolution approach and effective collisionless boundary conditions yield the same plasma parameters profiles, from Ref. [4]. Verification of our model by comparison to modeling results from Ref. [5] is shown as well; exact agreement was obtained.*

---

[1] Current affiliation: Lawrence Livermore National Laboratory (LLNL). LLNL is operated by Lawrence Livermore National Security, LLC, for the U.S. Department of Energy, National Nuclear Security Administration under Contract DE-AC52-07NA27344

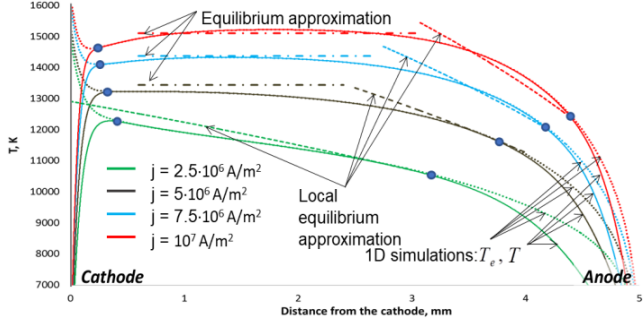

*Fig.2 Profiles of electron (Te) and heavy particles (T) temperature in the argon arc from Ref. [3] for various current densities, j. Uniform profile in the arc core (equilibrium region) corresponds to a balance of the Joule heating and radiation losses, $j^2 \rho_{el}(T_{eq}) = Q^{rad}(T_{eq})$. Here, $\rho_{el}$ is electrical resistivity, $T_{eq}$ is temperature of the equilibrium, $Q^{rad}$ is volumetric radiative heat loss. Closer to the anode, temperature profiles are non-uniform, though thermal equilibrium remains (local equilibrium region). Non-uniform profile in this region can be determined from energy balance neglecting radiation, thermal conduction and the electric field; the electric current is driven by electron diffusion $\vec{j} = \frac{kT}{e\rho_{el}}\left(\frac{\nabla n_e}{n_e} - 1.5\frac{\nabla T}{T}\right)$. Assuming Saha relation in this region, yields $E_{ion}T^{1.5}\nabla T = 2\tilde{\rho}\vec{j}$. Here, k is the Boltzmann constant, e is elementary charge, $E_{ion}$ is ionization energy, $\tilde{\rho}$ is electrical conductivity factor (see Ref. [2] for details). In the vicinity of the electrodes, thermal equilibrium is violated.*

For simulations of arc plasma properties and particle transport we consider a mixture of neutrals, ions and electrons. A system of the fluid equations in two dimensions was solved for the mixture velocity, and density. Separate energy equations were solved for heavy particles and electrons. Diffusion equations were solved for all heavy particles except background gas. Concentration of the background gas was obtained algebraically from the mixture and other particles densities. Electron transport equation resulted into equation for the electric potential[5]. The gas mixture and electron temperature profiles are simulated taking into account radiation and heating of the anode and cathode. Cooling of the anode due to the evaporated flux of carbon atoms and molecules, heating of the cathode due to deposition of carbon atoms and thermal resistance of carbon deposit are all taken into account in the heat balance as well as other surface processes including electron emission, ion recombination, radiation and effects of space-charge sheaths. Electron diffusion is also important to take into account and is responsible for significant increase of the electric field near the cathode and may also lead to the electric field reversal near the anode. We therefore needed to use a code that can model multiple fluids, neutral and charged, and the heat exchange and radiation to properly account for the intense heat transfer in the arc, as well as conjugate heat transfer in plasma and solid domains.

(a) Schematic of processes in the cathode region:

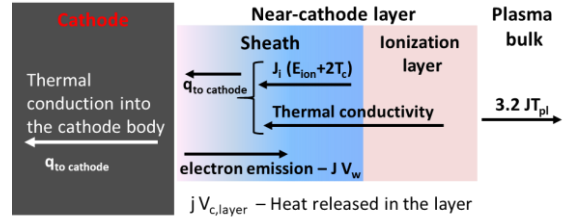

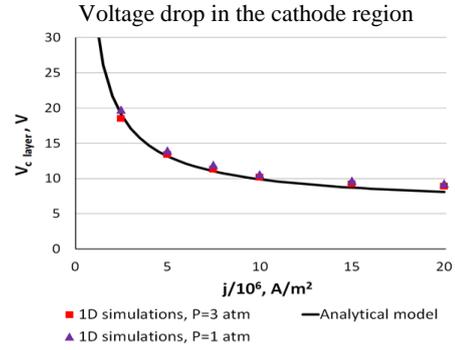

(b) Ion current to the cathode:

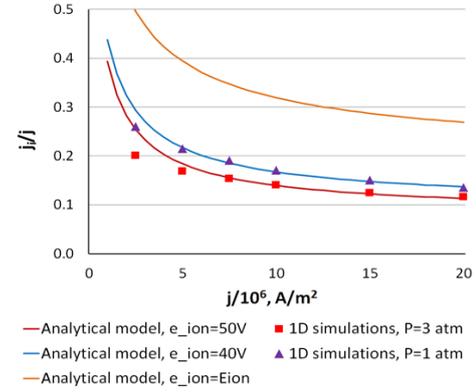

(c) Width of the cathode region:

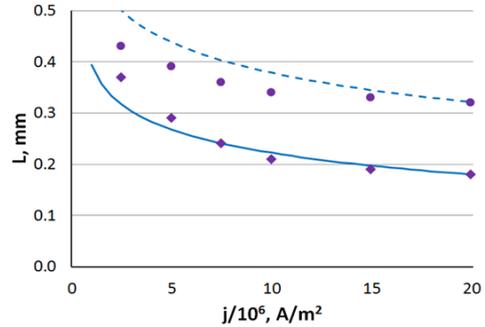

*Fig.3 Results of analytical model of the cathode region from Ref.[3]. Cathode voltage shown in Fig.3b is determined from balance of heat fluxes shown in Fig.3a, $jV_{c,layer} = 3.2jT_{pl} + jV_w + q_{cathode}$, where $T_{pl}$ is*

*temperature of the plasma, $V_w$ is work function of the anode material, $q_{cathode}$ is heat flux in the cathode; and the ion current $j_i$ is determined from the heat flux to the cathode, $q_{to\ cathode} = \varepsilon_{ion} j_i$, where $\varepsilon_{ion}$= 50 V is the ionization cost Fig.3c; (d) width of the cathode region is determined from the ion diffusion equation $\frac{d}{dx}\left(D\frac{dn_e}{dx}\right) = k_r n_e^3 - k_i n_a n_e$, where $k_r$ and $k_i$ are ionization and recombination coefficients.*

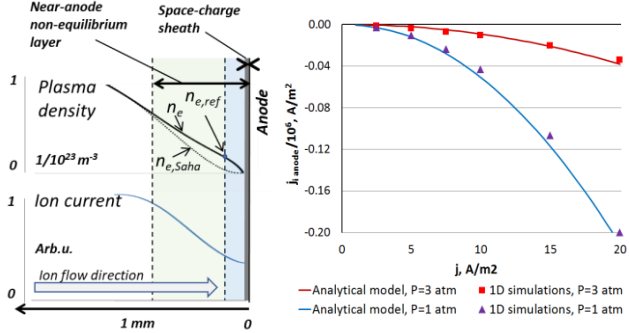

*Fig.4 Anodic region. Analytical solution for the plasma density $\left(\frac{n}{n_{Saha}}\right)^2 \approx 1 + \frac{C}{k_i(T) p T^{2.5}}$ and ion current to the anode $\Gamma_{i,a} \approx AB \frac{n_{ref}}{T_{ref}} \frac{2eE_{ion} - kT_i + 4kT_{ref}}{2eE_{ion} - kT_i - 6kT_{ref}}$ from Ref. [3].*

ANSYS-CFX is a robust Computational Fluid Dynamics (CFD) code allowing 3D simulations in areas of complex geometry using unstructured grids and multiple processors. The code incorporates models for various physical phenomena, including the heat and mass transfer, radiation transfer, buoyancy effects, etc. The code is extendable – it allows implementation of additional processes by introducing additional variables and transport equations with the boundary conditions. Solution with CFX can be obtained simultaneously in gas/plasma and solid domains, allowing conjugate heat transfer simulations.

ANSYS-CFX code was significantly customized in order to implement an a model of a carbon arc in helium atmosphere for synthesis of nanoparticles. The self-consistent arc model was developed and consisted of the Navier-Stokes equations for the plasma and gas phase coupled with models for the heat transfer and current flow insolid electrodes, for the ablation of the anode and deposition at the cathode, as well as the model for near-electrode space-charge sheaths introduced as wall function type boundary conditions, introduced at the plasma-electrode interfaces. Model also takes into account the radiation from electrodes surfaces, Joule heating of electrodes, and the thermal resistance of the deposit at the cathode. Note that radiation from electrodes appeared to be a large fraction of the heat flux.

Accurately accounting for the near-electrode space-charge sheaths required consideration of nonlinear coupling in the sheath regions of the electron plasma current, electron emission current, ion current, sheath voltage drop, heat fluxes, ablation/deposition. Significant effort was required to elaborate the numerical procedure for coupling of so many different plasma variables at the plasma-electrode interfaces and its implementation into ANSYS CFX code.

Full set of transport equations for plasma/gas region comprises the following equations: Navier-Stockes equations for the mixture including gravity:
$$\nabla \cdot (\rho \vec{v}\vec{v}) = -\nabla p + \nabla \cdot (\mu \nabla \vec{v}) + \rho \vec{g} \quad (1)$$
$$\nabla \cdot (\rho \vec{v}) = 0 \quad (2)$$

Here, $\rho$ is the gas density, $\vec{v}$ is mean flow velocity, $p$ is pressure, $\mu$ is viscosity and $\vec{g}$ is gravitational acceleration.

Neutral transport equation for carbon atoms in helium background gas:
$$\nabla \cdot (\rho c_C \vec{v}) = \nabla (D\nabla(\rho c_C)) - S_i \quad (3)$$

Here, $D$ is diffusion coefficient and $c_C$ is mass fraction of carbon gas.

Ion transport equation including thermal diffusion effects and friction with electrons:
$$\nabla \cdot (n_i \vec{v}) = \nabla(D_a \nabla n_i + D_{thdiff} \nabla T + D_{thdiff,e} \nabla T_e + j_e \gamma_{e,i}) + S_i$$
$$S_i = \alpha n_e n_{C,Ar} - \beta n_e^3 \quad (4)$$

Here, $D_a$ is the ambipolar diffusion coefficient, $D_{thdiff}$ is thermal diffusion coefficient, $S_i$ is net ionization rate.

Electron transport equation including thermal diffusion effects (effects of drift, diffusion and thermal diffusion) (this equation can be used to determine the electric potential $V$ and electric field $\vec{E}$):
$$\vec{E} = -\nabla V = -\frac{k}{e}\left(1 + C_e^{(e)}\right)\nabla T_e - \frac{k}{e}T_e \nabla \ln n_e + \frac{\vec{j}_e}{\sigma}. \quad (5)$$

Electric potential is found from charge conservation: $\nabla \vec{j} = 0$.

Equation of state: $p = (n_{neutrals} + n_i)kT + n_e kT_e$. (6)

Electron energy balance equation:
$$\nabla \cdot \left((2.5 + A_e)kT_e \frac{\vec{j}_e}{e}\right) =$$
$$\nabla \cdot (\lambda_e \nabla T_e) - S_i E_i - Q^{electrons-heavy} - Q^{rad} + \vec{j} \cdot \vec{E} \quad (7)$$

Heavy particles energy balance equation:
$$\nabla \cdot (\rho \varepsilon \vec{v}) = \nabla \cdot (\lambda \nabla(T)) + Q^{electrons-heavy} + \vec{j}_i \cdot \vec{E}. \quad (8)$$

Quasi neutrality relation: $n_e = n_i$. (9)

Here, $Q^{electrons-heavy}$, $\sigma$, $\lambda_e$ are functions of $(T_e, T, n_e, n_a, Q_{e,i}, Q_{e,a})$, $Q^{rad} = f(T_e, p)$, and expressions for other transport coefficients see Refs.[3,4] for details.

To account for the ablation and deposition, a new boundary condition for ablation flux, $G_{abl}$, was developed enabling automatic selection between these opposite processes

$$G_{abl} = \left(p_{satur,C}(T) - p_C\right)\sqrt{M_c/(2\pi RT)}. \quad (10)$$

Here, $p_C = n_C kT$ is partial pressure of carbon in the vicinity of an electrode surface, and $p_{satur,C}(T)$ is the saturated pressure of carbon vapor, T is temperature of an electrode. Equation (8) was used in the same form on cathode and anode surfaces, so ablation and deposition rates were determined by the solution at the interfaces, and are not defined *a priori*. Equation (8) was included in variable mass, momentum and energy source on the interface.

At chamber walls, we used zero current condition (non-conducting walls) and the no-slip condition for the fluid velocity, $\vec{v}=0$. For current and voltage boundary conditions on the electrodes, we took into account effect of electron emission on sheath formation and current generation, see Fig. 5 and Refs.[3,4] for details:

$$V_a = V_{plasma,a} + \Delta V_{sh,a}, \quad V_c = V_{plasma,c} + \Delta V_{sh,c} \quad (11)$$

$$j_{n,plasma} = j_e^{emission} - j_e^{plasma\ to\ electrode} + j_i = f(\Delta V_{sh}, T, T_e, n_e),$$
$$j_{n,plasma} = j_{n,electrode}. \quad (12)$$

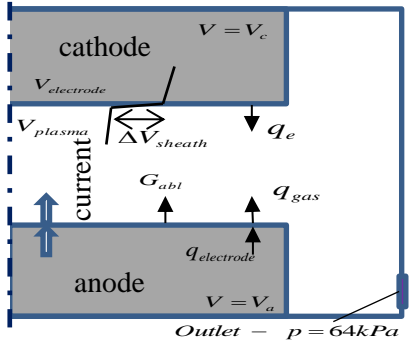

*Fig. 5. Schematics of boundary and interfacial conditions.*

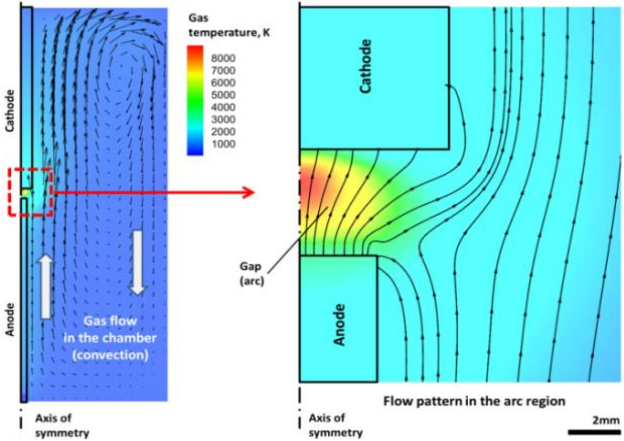

*Fig. 6. Results of simulations: temperature profile and flow pattern in the whole chamber (a) and zoom-in the arc region (b).*

## 2. Simulations results

Comprehensive parametric simulation study was performed for the carbon arc in helium atmosphere (500 torr, which corresponds to most of experimental conditions) for various arc currents, electrodes radii and inter-electrode gap sizes. The simulations were 2D-axisymmetric and steady-state. Simulation results were used for interpretation of the experimental data and as an input for models of diagnostics of nanoparticles in Refs. [6,7].

Simulations of flow pattern have shown that taking into account of the carbon deposition at the cathode strongly affects the gas flow, making it almost straight from anode to cathode, as shown in Fig. 6, opposite to previous assumptions[1], and strongly affects the carbon density and velocity profiles, especially in the region of nanoparticle synthesis.

(a) Ablation and deposition rates as a function of arc current:

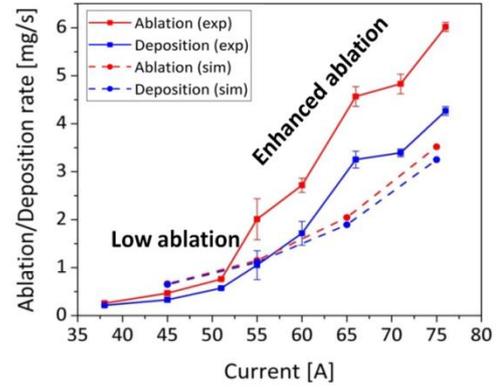

(b) Ablation and deposition rates as a function of inter-electrode gap width:

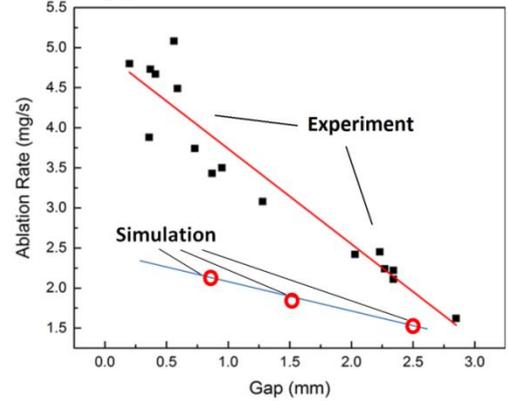

*Fig.7 Ablation (a, b) and deposition (b) rates as a function of arc current (a) and inter-electrode gap width (b). Cathode diameter was 9mm, anode diameter was 6mm.*

For validation of the arc model and benchmarking of the customized code, results of 2D simulations were compared with available experimental data. In the cases considered, cathode diameter was 9mm, anode diameter was 6mm (unless stated otherwise). One of the important values characterizing the synthesis process in the carbon arc is the anode ablation rate. Figure 7 displays variation of the ablation rate as a function of the arc current (inter-electrode gap size is 1.5 mm) and as a function of inter-electrode gap width (the arc current is 60 A). Quantitative agreement between the simulation results and experimental data is observed in case (a) and qualitative agreement in case (b). In Figure 8(a), the radial profile of plasma density in the middle cross-section of the arc is shown. Figure 8(b) displays the total arc voltage as a function of the anode diameter, both showing good agreement. The reasons for some differences between the simulation results and experimental data are being investigated.

(a)  Plasma density:

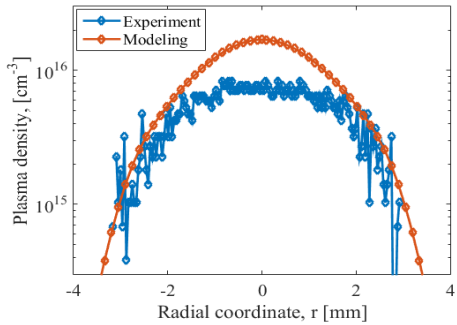

(b)  Arc total voltage VS anode diameter:

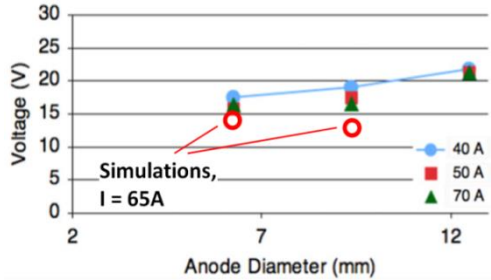

*Fig.8. Plasma density profile (a) and the arc voltage as a function of anode diameter (simulations are empty red circles) (b). Experimental data shown by filled symbols in Fig.8 (b) are taken from Ref. [8].*

Current in the arc flows in the central region of the arc and is presented in Fig.9. Figure 9(a) displays the simulated profile of the electric potential, which appears to be non-uniform along the cathode surface due to large variations in the cathode sheath voltage drop. This non-uniformity of current density leads to constriction of the arc plasma near the cathode and formation of a cathode spot (Fig.9 (b)); simulation results are in accordance with the experimental observation shown in Fig.9(c), where emission of $H_\alpha$ spectral lines shows hot core of the arc (small amount of hydrogen, 5%, was added to helium for this experiment). Experimental study of the width of arc current flow channel near the cathode was performed in Ref. [9]. In the experiment, the cathode consisted of 2 segments: the central segment of diameter 3.2 mm and the periphery. Almost all of the current (95%) was observed to be flowing through the central segment. Simulations give a comparable result: 75% of current flows through a circle of diameter 3.2 mm. The difference can be explained by the fact that segmented cathode did not allow radial heat transfer through its body limiting the cathode spot size.

Electric current streamlines:

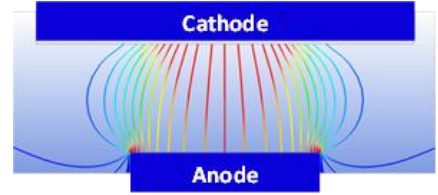

Electric potential profile:

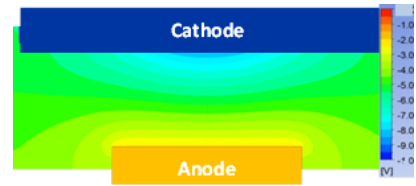

Experimental data from emission of $H_\alpha$ spectral lines, Ref. [6]:

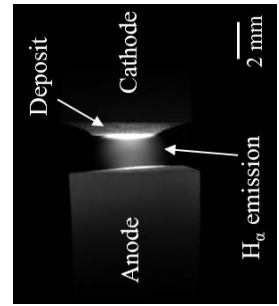

*Fig.9. Computational and experimental results showing constriction of the arc current flow channel near the cathode.*

In simulations of carbon dimers we used thermodynamic approach, see Ref. [10] for details. Simulated spatial profile of carbon dimers ($C_2$ molecules) density is shown in Fig.10 and is very close to the experimental profile[10]. Note that good prediction of the spatial density profile of carbon dimers is a crucial result, because it means that processes of carbon ablation, deposition and convection/diffusion, heat production and transfer are well reproduced by the model. Therefore, the model has predictive power and can be used to determine conditions (regions) of nanoparticle synthesis.

We also developed analytical model for the transition of the arc operation from low-ablation mode to high-ablation mode observed in our previous experiments[11]. The model can be used to self-consistently determine the profiles of the electric field, electron density and electron temperature in the near-anode region of the arc discharge. Simulations of the carbon arc predicted that in the low-ablation mode, the arc current to anode is driven mainly by the electron diffusion to the anode[12]. Reference [13] proposed theoretical explanation for the transition from high-ablation mode to low-ablation mode as follows. In the high-ablation mode the anode sheath voltage is high (positive anode sheath), close to the ionization potential of anode material, therefore providing high heat flux and high ablation flux from the anode. While for the low-ablation mode the anode voltage is low (negative anode sheath) an order of magnitude smaller than the ionization potential of anode material; hence the heat flux to the anode is small and ablation rate is low.

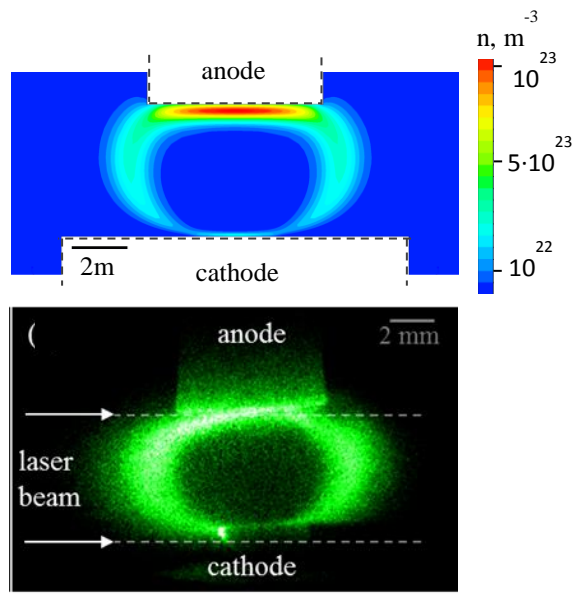

*Fig.10 Carbon dimer distribution has a bubble-like shape around the arc core. Comparison of (top) simulated and (bottom) measured $C_2$ density distribution (planar LIF: spectral image of carbon dimer, emission is at 470 nm and laser radiation is at 437 nm) in the arc at current 50 A; see Ref. [6] for details.*

### 3. Summary

The developed mathematical model of the short arc was used for prediction of nanomaterial synthesis. The model did not use simplifying assumptions as previous models, such as the local thermodynamic equilibrium (LTE) (the same temperature for the electrons and heavy species), and the ionization and recombination balance (the Saha equation). The new model also accounts for the heat losses in ablation processes and heat generation in carbon deposition processes; as well as space-charge sheaths near the electrodes[12]. The model was implemented into the commercial 3D code ANSYS CFX.

With the newly developed code, comprehensive parametric simulation study of the carbon arc in helium atmosphere was performed for various arc currents, diameters of the electrodes and inter-electrode gap sizes. The simulation results were used for interpretation of the experimental data[6,7].

For benchmarking of the arc model, an additional 1D code that resolves the sheath regions was written and simulations were performed to compare with results of previous numerical studies,[5] and complete agreement was achieved[4]. We also developed analytical arc model that can readily predict the heat fluxes to the electrodes, near-electrode voltage drops, total arc voltage, temperature profile, etc[3].

The arc modeling was supported by the US DOE Office of Science, Fusion Energy Sciences. Experiments and simulations of synthesis processes were supported by the US Department of Energy (DOE), Office of Science, Basic Energy Sciences, Materials Sciences and Engineering Division.